\documentstyle[aps,preprint,epsf,floats]{revtex}
\begin{document}
\draft
\title{Impurity effect on low-temperature polarisation of the
charge-density-waves in o-TaS$_3$}
\author{N.~I.~Baklanov, S.~V.~Zaitsev-Zotov, and V.~V.~Frolov}
\address{Institute of Radioengineering and Electronics of Russian 
Academy of Sciences, \\
Mokhovaya 11, 103907 Moscow, Russia. \\}
\author{P.~Monceau}
\address{Centre de Recherches sur Les Tr\`{e}s Basses Temp\'{e}ratures,\\
C.N.R.S., B.P. 166, 38042 Grenoble C\'{e}dex 9, France\\}
\preprint{Accepted for publication in Physics Letters A}
\maketitle
\begin{abstract}
The temperature dependence of the low-temperature dielectric response is 
studied in o-TaS$_3$ samples doped by Nb, Se, and Ni and for nominally pure 
ones. It is found, that the low-temperature dielectric constant depends  
anomalously on doping and is higher for doped crystals, whereas the 
temperature dependence of the characteristic time of all samples follows 
the activation law with nearly the same 
activation energy $\sim~400$~K ($T>20$~K). The observed behaviour is 
inconsistent with all available explanations of the low-temperature 
dielectric anomaly.\\
\end{abstract}

\section{Introduction}

Quasi-one dimensional conductors like TaS$_{3}$, K$_{0.3}$MoO$_{3}$ {\it etc.} 
are well known physical systems which undergo spontaneous structural
transition into a low-temperature ordered state, where one of the most
widely studied electronic crystal - the charge-density waves (CDWs) - is
formed \cite{ECRYS}. Similarity between the physical properties of the CDW
conductors, Wigner crystals, vortex lattices in superconductors, together
with the specific possibility of measurements of extremely small
displacements of the electronic crystal, that in the case of the CDW is
associated with extremely high dielectric constant $\varepsilon $ up to 
$10^{8}$, makes the CDW conductors one of the most convenient model system
for investigation of the collective response to an external force.

In quasi-one dimensional conductors at an electric field high enough to
overcome the CDW-impurity interaction the CDW starts to slide providing a
contribution in nonlinear conduction. At relatively small electric fields
the CDW is pinned by impurities, and its response to an electric field may
provide an information on CDW-impurity interaction, as well as on the CDW
state itself. In the recent years the interest to dielectric properties of
the CDW conductors have been revived because of observation of the
low-temperature anomaly of their $\epsilon (\omega ,T)$ for low frequencies 
$\omega \sim 10^{-2}-10^{5}$~Hz \cite{Kriza,KrizaKim,Ong,NadSSC,NadPRB}. The
origin of this anomaly remains to be the subject of intensive study and
discussion \cite{Kriza,KrizaKim,Ong,NadSSC,NadPRB,Volkov,Larkin}.

Many experimental facts show that the CDW at sufficiently low temperatures
is characterised by ''glassy'' behaviour (see \cite{Bil-1} and references
therein) related to the existence of metastable states in the CDW system.
In particular, the rough free energy landscape caused by pinning of the CDW
leads to a continuous relaxation times spectrum that reveals itself in a
long-time logarithmic relaxation of polarisation. In the framework of this
``glassy'' approach the temperature maximum in $\epsilon (\omega ,T)$ marks
the critical slowing down of the CDW kinetics, when the system goes through
a temperature-dependent characteristic time $\tau (T)$ into a glassy state 
\cite{Kriza}, which in the case of TaS$_{3}$ was recognised as a 
disorder-induced glassy state~\cite{NadSSC,NadPRB}.

Alternative explanations of the low-temperature anomaly were given recently 
\cite{Volkov,Larkin}. On the first hand, Volkov~\cite{Volkov} analysed the
screening effects on a deformed CDW presented in the form of a periodic
chain of phase solitons pinned by impurities. It was shown, that the
characteristic time of the dielectric response of such a system follows an
activation law. In the framework of this approach the activation energy of 
$\tau (T)$ corresponds to the activation energy of the linear conduction, in
agreement with the experimental data~for K$_{0.3}$MoO$_{3}$ \cite
{Kriza,KrizaKim,Ong}, (TaSe$_{4}$)$_{2}$I \cite{Cava} and TaS$_3$ at 
$T\sim 100$~K~\cite{CavaFleming,Katica-tbp}. Note that 
for these materials, as well as for all known CDW materials, the dielectric 
constant {\it decreases} with increasing impurity concentration.

On the other hand, Brazovskii and Larkin \cite{Larkin} considered a
phenomenological model of pinning, taking into account both local and
collective types of pinning. It was shown that the characteristic time of
the dielectric response of the pinned CDW follows an activation (in the case
of a single pinning energy) or close to activation law (in the case of a
distribution of the pinning energy) with the activation energy close to the
pinning energy. The dielectric constant also {\it decreases} with increasing
impurity concentration.

So the question on the origin of the low-temperature CDW polarisation
anomaly remains to be open. At the same time, there is a way to distinguish
experimentally between the explanations given above. For this purpose it is
necessary to study the characteristic time, $\tau $, and the dielectric
constant, $\epsilon $, as a function of the temperature, impurity
concentration, $n_{i}$, and their chemical type. If the maximum of the
dielectric response is due to the disorder-induced glass 
transition~\cite{NadSSC,NadPRB}, the characteristic time $\tau $
increases with increasing impurity concentration $n_{i}$. If the maximum is
related to the screening mechanism~\cite{Volkov}, $\tau (T)$ has the same
activation energy as the screening carrier concentration (i.e. as the linear
conduction). If the maximum is related to the pinning energy, 
the activation energy depends on a chemical
type of dominating doping impurity, but is independent of their
concentration. In all cases 
\cite{Kriza,KrizaKim,Ong,NadSSC,NadPRB,Volkov,Larkin} the dielectric constant 
{\it decreases} with increase of impurity concentration. 

In this paper we present results of the real-time low-temperature
polarisation measurements of o-TaS$_{3}$ doped by Nb, Ni, and Se, as well as
for pure one. The measurements were carried out over the time interval from 
$1~\mu $s to $10^{3}$~s in the temperature range 1.6-32~K. Our results show
that the characteristic time follows the activation law with the activation
energy independent of the doping element and its concentration, whereas the
activation energy of the linear conduction varies within a factor 2. It is also
found that the dielectric constant is higher for impure crystals. Our results are 
therefore inconsistent with all available explanations of the low-temperature 
dielectric anomaly and indicate that there are subgap excitations of the CDW 
mediating its low-temperature polarisation.

\section{Experimental}

A real-time relaxation technique is appropriate to reveal the
characteristics of the transport properties of the CDW and the low-frequency
CDW excitations~\cite{Kriza,MihTess,Bakl}. The time-domain measurements are
free from the integral effects inherent to the frequency-domain technique in
case of presence of internal interactions in the system~\cite{book_diel}. We
studied the dielectric response of o-TaS$_{3}$ samples that arises during
the application of an alternating-sign periodic train of voltage pulses (see
inset in Fig.~\ref{curv1}). Unlike the earlier measurements on the blue
bronzes~\cite{Kriza,MihTess} the polarisation charge $Q$ was measured as a
function of time $t$ after switching the polarisation voltage across the
sample, $V$, to its {\em zero} value. The $Q(t)$ curves were averaged over a
large number of pulses (order of 100). Such zero voltage measurements allows
us to get rid of the ohmic contribution of thermally activated
quasi-particles, that is essential at long times and high temperatures. For
the measurements we have used a hand-made integrator on the base of LMC6001
ultra-low input current amplifier. The low-frequency noise of the apparatus
is of the order of $3\times 10^{-16}$~A/$\sqrt{Hz}$ (at 1 kHz).

In addition, to compare the time-domain technique with frequency-domain 
technique we undertook measurements of real and imaginary parts of AC 
conduction in a frequency range 
10 Hz - 10 kHz with using SR530 lock-in amplifier as a phase-sensitive detector.

o-Ta$_{3}$ crystals as well as bundles originating from two different
sources were investigated. Three batches were prepared at the Switzerland
University and was kindly supplied to us by Dr. F.~Levy. Other ones were
prepared by us in IRE RAS (Moscow, Russia). The contacts to the samples were
cold soldered 50-$\mu $m In wires. The samples characteristics are briefly
summarised in Table~\ref{table1}.

\begin{table}
\caption{}
\begin{tabular}{|c|c|c|c|c|c|c|c|}
Sample & Nominal  & $T_{p}$ & $E_{T}$ & 
$E_{\sigma }$ & $E_{\epsilon }$ & $\epsilon (0.1{\rm~s} )^{\dagger }$ 
& $\sigma^{\prime\prime} ({\rm 1~kHz,10~K})/\sigma ({\rm 100~K})$ \\ 
& Impurity Content &    K & V/cm & K & K & $T = 20$~K &\\ 
\hline
1$^{a,b}$ & pure &  221 & 0.4 & 400 & 420 & $1.33\times 10^{6}$&$3.9\times 10^{-5}$ \\ 
2$^{a,b}$ & pure &  221 & 0.4 &  & 456 & $1.76\times 10^{6}$& \\ 
3$^{c}$ & pure &   220 & 1.4 & 360 & 437 & $3.43\times 10^{6}$& $1.8\times 10^{-5}$\\ 
4$^{c}$ & 0.2\%Nb  & 205 & $>35$ & 370 & 440 & $3.31\times 10^{6}$& $1.5\times 10^{-4}$\\ 
5$^{c}$ & 0.5\%Nb  & 202 & $>20$ &  & 476 & $7.15\times 10^{7}$& $6.1\times 10^{-4}$\\ 
6$^{b}$ & 5\%Ni  & 219 & $^{\ddagger }$ & 370 & 395 & $1.14\times 10^{7}$& $9.9\times 10^{-5}$\\ 
7$^{b}$ & 1\%Se  & 212 & 5 & 260 & 446 & $1.81\times 10^{7}$ &
$1.9\times 10^{-4}$\\ 
\end{tabular}
$^{a}$Samples from the same batch.\\
$^{b}$Samples prepared in IRE\ RAS (Moscow, Russia).\\
$^{c}$Samples prepared in the Institut de Physique Appliquee (Lausanne,
Switzerland).\\
$\dagger $The dielectric constant was calculated as $\epsilon
(t)=(1/\epsilon_0)[Q(t)/V](L/S)$ where $L$ is the length and $s$ is the
cross-sectional area of a sample. \\
$^{\ddagger }$I-V curves are nonlinear and have no threshold.
\label{table1}
\end{table}

\section{Results}

First of all we got sets of polarisation charge {\it vs.} polarisation pulse
amplitude dependencies to find the linear response region. A voltage
corresponding to this region for all temperatures was chosen for further
detailed study of the response over entire temperature range. The results
presented below corresponds to the linear response region.

The curves obtained at a fixed polarisation pulse duration $t_p$ are well
reproducible. Comparing the $Q(t)$ curves measured for various $t_p$ we have
detected a weak dependence of $Q(t)$ on different $t_p$, like it was
observed recently in K$_{0.3}$MoO$_3$~\cite{history}. It has been checked up
that the results reported here are not affected by this observed dependence
and correspond to the infinite duration of the polarisation pulse.

\begin{figure}
\vskip -2cm
\epsfysize=17cm
\centerline{\epsffile{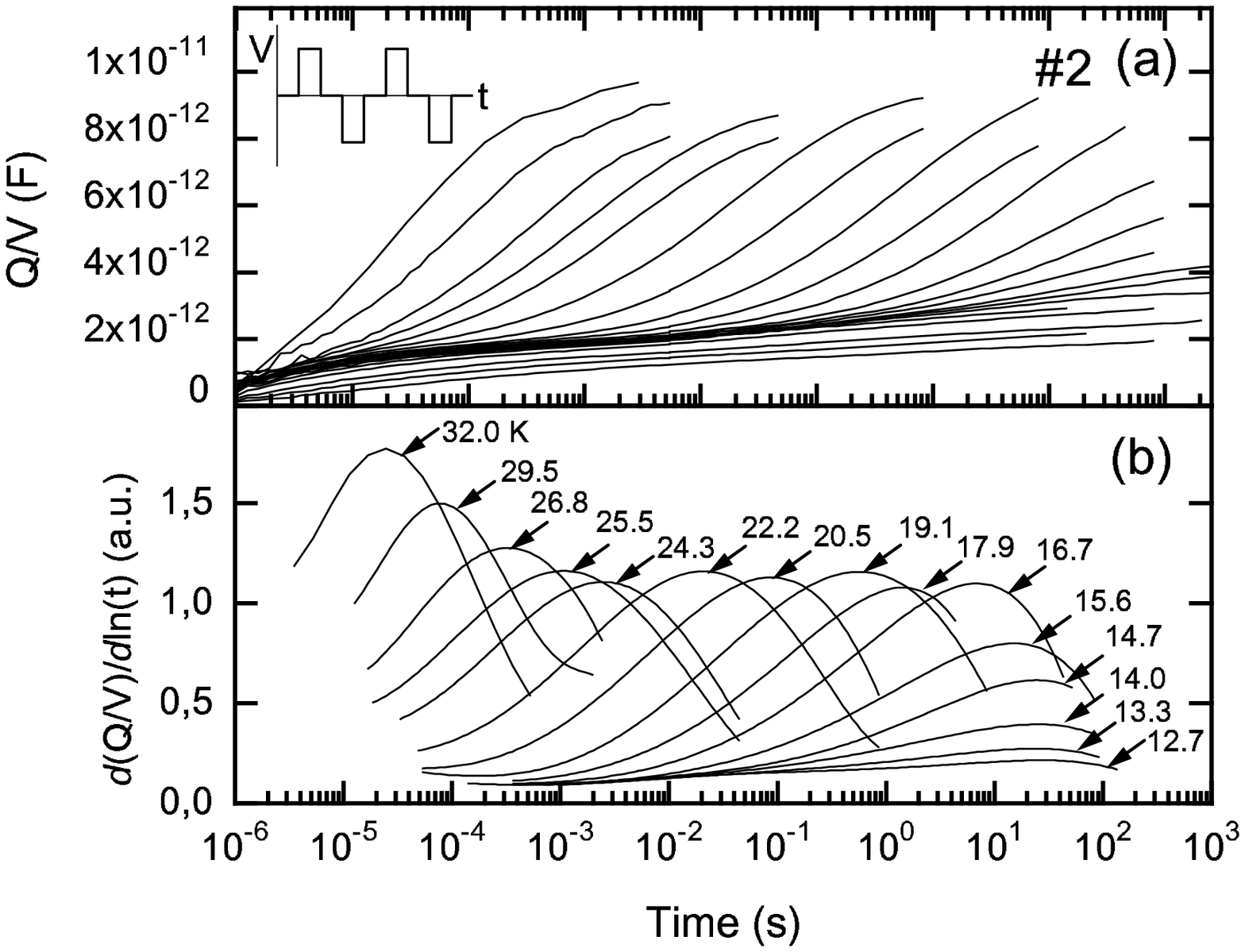}}
\vskip -6cm
\caption{(a) Typical real-time relaxation of the polarisation charge curve
set of o-TaS$_{3}$ crystal (\#2) in the linear response region. (b)
Respective set of the logarithmic derivatives for the same sample.}
\label{curv1}
\end{figure}

Fig.~\ref{curv1}(a) demonstrates a typical set of $Q(t)$ curves for pure and
doped TaS$_{3}$ samples. The relaxation of the polarisation is
nonexponential; it continues to the longest times at which measurements were
carried out.

For all samples at temperatures above approximately $T=10$~K we have observe
a region with a comparatively rapid growth of $Q(t)$ with time. 
This relaxation mode gives a sharp maximum on a logarithmic derivative 
$dQ/d\log t$ of the polarisation (Fig.~\ref{curv1}(b)) indicating passing of
the temperature-dependent characteristic time; in agreement with earlier
studies \cite{Kriza,KrizaKim,Ong,NadSSC,NadPRB,Bakl}.

Fig.~\ref{curv1}(b) shows the respective set of the logarithmic derivative
of the relaxation curves, $d(Q/V)/d\log t$. The characteristic time, $\tau$, 
is determined as a time corresponding to the maximum of the derivatives.

\begin{figure}
\vskip -2cm
\epsfysize=14cm
\centerline{\epsffile{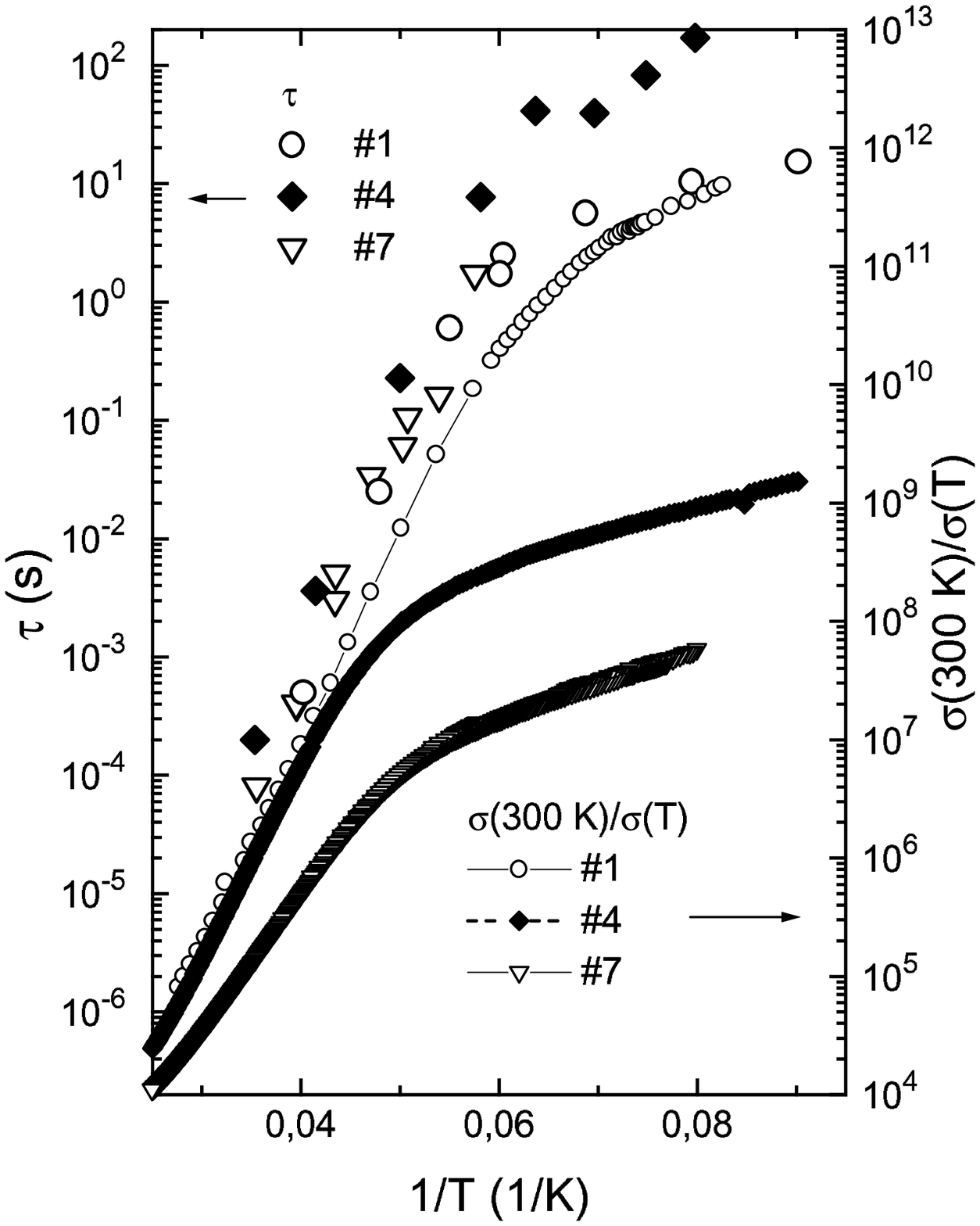}}
\vskip -2cm
\caption{Temperature-dependent characteristic time found from the real-time
relaxation data together with the linear conduction data. }
\label{akt}
\end{figure}

Fig.~\ref{akt} shows typical temperature dependencies of the characteristic
time $\tau $ of a pure and two doped samples. The high-temperature part of 
$\tau (1/T)$ dependence ($T\geq T^{*}\simeq 17$ K) can be fitted by the
activation law with the activation energies $440\pm 40$~K. The values of the 
crossover temperature $T^{*}$, as well as the activation energies 
$E_{\varepsilon }$, 
are remarkably nearly the same for all samples (within 10\% experimental error),
i.e. they are independent of a crystal purity (see Table~\ref{table1}). Moreover, the
relaxation time at a fixed temperature does not correlate to neither type
nor nominal concentration of the impurities in the studied samples. For
example, doping by Nb increases $\tau (20$ K$)$ from $8\times 10^{-2}$~s
(sample 3) to $2\times 10^{-1}$~s (sample 4) and $5\times 10^{-1}$~s (sample
5), whereas doping by Ni decreases it down to $2\times 10^{-2}$~s (sample 6)
with respect to the most pure crystals 1-3.

We  also observed another branch of $\tau (1/T)$ dependence. Note
that the heights of maximums of $d(Q/V)/d\log t$ curves (Fig.~\ref{curv1}(b)) 
are approximately the same at $25\geq T\geq T^{*}$~K, but decrease
rapidly at $T<T^{*}$~K. Similarly, the slope of $\log \tau $ {\em vs.} $1/T$
curve at $T<T^{*}$ is roughly by factor 4 lower that at $T>T^{*}$. This
indicates that there is a change of the relaxation mechanisms around $T^{*}$. 
Suggesting the activation law for the low-temperature part of $\tau (T)$
one gets the estimate for the activation energy around $120\pm 50$~K
depending on doping impurity.

In addition, measurements of the DC linear conduction
temperature variation $\sigma (T)$ were carried out for the same samples for
the temperature range $T<50$~K. Fragments of $\log (1/\sigma (T))$ {\it vs.} 
$1/T$ curves are shown in Fig.~\ref{akt}.
A region of the activation dependence, $\sigma \propto \exp (-E_{\sigma }/T)$, 
is observed above 20 K, in agreement with the results of earlier studies.
Note that while the slopes of $\log (\tau(1/T))$ curves
are very close to each other, the slopes of $\log(\sigma (1/T))$ 
curves are different.
The respective set of $E_{\sigma }$ demonstrates their distinction from 
$E_{\epsilon }$ (See Table~\ref{table1}).

Direct measurements of the dielectric constant has shown that $\epsilon$ 
of impure crystals is higher than of pure ones (see Table~\ref{table1}).  
To verify this result we also measured the ratio 
$\sigma^{\prime\prime}(f,10{\rm~K})/\sigma ({\rm 100~K})$ 
($\sigma^{\prime\prime}$ 
is an imaginary part of the AC conduction) 
which is independent of the geometrical factors. 
As $\sigma ({\rm 100~K})$, is practically 
independent of impurity concentration, this ratio can be considered as a 
measure of the low-temperature dielectric constant. And again, it was found 
that $\sigma^{\prime\prime} ({\rm 1~kHz,10~K})/\sigma ({\rm 100~K})$ is 
higher for impure crystals than for pure ones. That reflects growth of the 
low-temperature $\epsilon$ with increase of impurity concentration. Note 
that this unusual low-temperature 
behaviour of $\epsilon(n_i)$ is opposite to the high-temperature behaviour 
corresponding to the relation $\epsilon \propto 1/E_{T}$ well-known for 
the high-temperature region \cite{eps}. 

\section{Discussion}

As was noted above, absence of any correlation between the characteristic
time and impurity content does not support the earlier 
suggestion~\cite{NadSSC,NadPRB} on 
the disorder-induced glass transition as a mechanism responsible for
the low-temperature dielectric anomaly.

In principle, the strong-pinning model~\cite{Larkin} may provide impurity
independent activation energy for the relaxation time. Namely, in the
original model~\cite{Larkin} the depinning barrier height depends on the
pinning potential and may be arbitrary large. One should expect, however,
that when the pinning potential exceeds some critical value, a new
alternative channel for depinning of the CDW appears. 
For example, depinning of the CDW
may occur through a suppression of the order parameter and local phase slip,
rather than through further shift of the continuous CDW. The respective
energy barrier depends mostly on parameters of the CDW itself
rather than on its interaction with impurity. Nevertheless, the
strong-pinning model~\cite{Larkin} predicts qualitatively different dependencies 
of the dielectric constant on the impurity concentration than it was observed.
Namely, it gives $\epsilon ^{-1}=\epsilon _w^{-1}+\epsilon _s^{-1}$,
where $\epsilon _w$ and $\epsilon _s$ are the weak and strong-pinning
components respectively. At $T<T^*$ both components are expected to
{\it decrease} with growing impurity concentration \cite{Larkin}, the value of 
$\epsilon$ should be lower for doped crystals. In contrast, our
experimental data clearly indicate growth of $\epsilon$
with increasing $n_{i}$ (see Table~\ref{table1}). So the strong-pinning 
model~\cite{Larkin} is inconsistent with the data. 

As was noted above, the screening model~\cite{Volkov} is consistent
with the high-temperature dielectric anomaly obseved in 
K$_{0.3}$MoO$_{3} $\cite{Kriza,KrizaKim,Ong}, (TaSe$_{4}$)$_{2}$I 
\cite{Cava} and TaS$_3$~\cite{CavaFleming,Katica-tbp} at $T\sim T_P/3$, 
where CDW deformations are screened by normal carriers exited over the 
Peierls gap. To apply similar explanation to the low-temperature anomaly in 
TaS$_3$ one needs to suggest existence of screening carriers with the 
activation energy around 400 K~\cite{solitons}. However, the present version of 
the screening theory~\cite{Volkov} can hardly account for the opposite 
relations between $\epsilon$ and $n_i$ for high- (decrease of $\epsilon$ 
with $n_i$) and low-temperature ranges (increase of $\epsilon$ with $n_i$).

So neither of available explanations of the low-temperature
dielectric anomaly of TaS$_3$ is consistent with the actual dependencies
of $\epsilon (n_i)$ and $\tau(T,n_i)$. The most surprising result is growth
of the dielectric constant with increase of impurity 
concentration (see Table~\ref{table1}). Note, such a behaviour may reflect the decrease of the 
pinning energy $W_p$ per the phase-correlation volume with growth of 
$n_i$: in the weak-pinning case $W_p\propto 1/n_i$ (see e.g. 
Ref.~\cite{Thorne} and references therein). Thus growths of both the 
low-temperature linear DC conduction observed earlier~\cite{PinEn} 
and AC conduction reported here with increase of the impurity concentration
are consistent with general expectations 
for creeping of weakly pinned CDW. Additional contribution to 
$\epsilon$ may also come from  increase of CDW dislocation density 
with increase of $n_i$~\cite{dislocations}. 

Another surprising result is independence of $E_{\epsilon}$ on impurity
type and concentration. This result indicates the existence 
of low-energy excitations in TaS$_3$ that mediate the relaxation process.
Though possible contribution of such excitations into low-temperature 
transport properties of TaS$_3$ was broadly discussed earlier~\cite{solitons}, 
no their contribution into the low-temperature conduction was 
observed in the present measurements.

In conclusion, the characteristic time of low-temperature polarisation of TaS$_3$
is found to follow the activation law with the activation energy $\sim 400$~K 
independent of the dominant impurity and its concentration. In addition, 
the low-temperature dielectric constant of doped samples is found to be 
higher then of pure ones. The observed behaviour is inconsistent with all
earlier suggested explanations of the low-temperature dielectric anomaly
and implies existence of low-energy excitations having the activation energy 
400~K and mediating the polarisation process. 

\section{Acknowledgments}

We are grateful to K.~Biljakovi\'{c}, S.~Brazovskii and A.F.~Volkov for many
useful discussions, and F.~Levy for supplying some samples of o-TaS$_{3}$.
This work was supported by RFBR grants 97-02-17751, 98-02-16667, 98-02-22061, 
by MNTP ''Physics of Solid-State Nanostructures'' (grant 97-1052), and by C.N.R.S. 
through jumelage 19 between C.R.T.B.T. and IRE RAS.


\begin{references}
\bibitem{dislocations}Note that in TaS$_3$ the threshold field for the onset
of sliding of the CDW growths exponentially with lowering temperature. So
one can expect that at sufficiently low temperatures nucleation of CDW 
dislocations reducing the CDW strain is energetically favourable. As a result, 
the CDW gets additional degree of freedom and may be more polarisable 
than a rigid CDW without dislocations.
\end{references}
\end{document}